\documentclass[aps,pre,twocolumn,showpacs,groupeaddress,preprintnumbers,floatfix]{revtex4-1}
\usepackage{graphicx,bm,amsmath,amssymb,natbib,url,epsfig}
\usepackage{dcolumn}
\usepackage{hyperref}
\usepackage{amsmath,amssymb}

\usepackage{color}
\definecolor{darkgreen}{rgb}{0.1,.6,.1}
\definecolor{greenblue}{rgb}{0.0,.1,.4}

\usepackage[normalem]{ulem}

\begin{document}

\title{Chimera patterns induced by distance-dependent power-law coupling in ecological networks} 

\author{Tanmoy Banerjee${}^{1}$}
\author{Partha Sharathi Dutta${}^{2}$}
\author{Anna Zakharova${}^{3}$}
\author{Eckehard Sch{\"{o}}ll${}^{3}$}
\affiliation{%
${}^1$Chaos and Complex Systems Research Laboratory, Department of Physics, University of Burdwan, Burdwan 713 104, West Bengal, India.\\
${}^{2}$Department of Mathematics, Indian Institute of Technology Ropar, Rupnagar 140 001, Punjab, India.\\
${}^3$Institut f{\"u}r Theoretische Physik, Technische Universit\"at Berlin, Hardenbergstra\ss{}e 36, 10623 Berlin, Germany.}%

\received{:to be included by reviewer}
\date{\today}

\begin{abstract}
This paper reports the occurrence of several chimera patterns and the associated transitions among them in a network of coupled oscillators, which are connected by a long range interaction that obeys a distance-dependent power law. This type of interaction is common in physics and biology and constitutes a general form of coupling scheme, where by tuning the power-law exponent of the long range interaction the coupling topology can be varied from local via nonlocal to global coupling. To explore the effect of the power-law coupling on collective dynamics, we consider a network consisting of a realistic ecological model of oscillating populations, namely the Rosenzweig--MacArthur model, and show that the variation of the power-law exponent mediates transitions between spatial synchrony and various chimera patterns. We map the possible spatiotemporal states and their scenarios that arise due to the interplay between the coupling strength and the power-law exponent. 
\end{abstract}

\pacs{05.45.Xt, 05.65.+b, 87.23.Cc}
\keywords{Amplitude chimera, chimera death, amplitude death, nonlocal coupling, ecological model}
\maketitle 

\section{Introduction}
\label{sec:intro}

The chimera state is an intriguing and counterintuitive spatiotemporal state that has been in the center of active research over the past decade \cite{chireview}. In this state the population of coupled {\it identical} oscillators spontaneously splits into two incongruous domains: In one domain the neighboring oscillators are synchronized, whereas in another domain the oscillators are desynchronized.  After its discovery in phase oscillators by Kuramoto and Battogtokh \cite{KuBa02} several theoretical studies \cite{st1,st2,scholl2} established the existence of this classical chimera state. Later on, other types of chimera states have also been discovered. Amplitude-mediated chimeras were reported in \cite{amc_sethia} where, under strong global coupling, incoherent fluctuations occur in both the phase and the amplitude in the incoherent domain. Recently, Zakharova {\em et al.}~\cite{scholl_CD} discovered pure {\it amplitude chimeras} where all the oscillators have the same phase velocity but they show uncorrelated fluctuations in amplitude in the incoherent domain. In this context several more general chimera patterns, like chimera death (CD) \cite{scholl_CD,tanCD} and chimeralike coexistence of synchronized oscillation and death (CSOD) \cite{csod} have recently been revealed. Chimera death (CD) is the steady state version of amplitude chimera, i.e., the population of oscillators in a network splits into incongruous coexisting domains of spatially coherent oscillation death (OD) and spatially incoherent OD. In the CSOD state, the population of oscillators split into two coexisting distinct, spatially separated domains: In one domain oscillators are oscillating coherently and in another domain neighboring oscillators randomly populate an oscillating state and a stable steady state.

Recent experimental observation of chimera states have established their robustness in natural and man-made systems. The first experimental observation of chimeras was reported in optical systems \cite{HaMuRo12} and chemical oscillators \cite{TiNkSh12}. Later, chimeras have been observed experimentally in  mechanical systems \cite{MaThFo13,KAP14}, electronic~\cite{LAR13,GAM14}, optoelectronic delayed-feedback \cite{LAR15} and electrochemical~\cite{WIC13,WIC14,SCH14a} oscillator systems, Boolean networks~\cite{ROS14a}, and optical combs \cite{VIK14}. Recently, chimera states have been observed in globally coupled networks of four optoelectronic oscillators \cite{raj_minimal}, similar to theoretical predictions for a small number of globally coupled lasers \cite{boe15}. In small networks chimeras are generally difficult to observe, but control methods to stabilize them have recently been proposed \cite{SIE14c,BIC15,OME16}. The strong current interest in chimeras may be attributed to their possible connection with several phenomena in nature, like unihemispheric sleep of dolphins and certain migratory birds \cite{st2,MaWaLi10}, ventricular fibrillation \cite{DaPeSa92}, and power grid networks \cite{grid1}. Recently, chimera patterns have been found in models from ecology \cite{csod,HIZ15}, SQUID metamaterials \cite{expt_meta}, and quantum systems \cite{schoell_qm} showing their omnipresence in the macroscopic as well as in the microscopic world.

In the studies on chimera, the coupling function always plays an important role. Initially it was believed that to induce chimeras a  nonlocal coupling is essential. Later it has been found that global coupling \cite{amc_sethia,SET14,YEL14,SCH14g,BOE15} and even local coupling \cite{csod,liang} may give rise to chimera states. In the nonlocal coupling, one has two control parameters: coupling range and coupling strength. The former is controlled by the kernel of the nonlocal coupling function: in previous studies generally a trigonometrically or exponentially decaying function or a rectangular kernel have been used. 

In this paper we focus on a more general and universal coupling scheme which is motivated by many real-world systems, but has not been investigated in the context of chimeras. It is governed by a long range interaction obeying a distance-dependent power law. In the long range coupling each node in the network is connected to all other nodes with an {\it effective interaction strength} that decreases with
increasing distance according to a {\it power-law}. Thus, the oscillators are subject to a long range interaction whose interaction strength is controlled by the {\it power-law exponent} (denoted by $s$). Earlier the long range interaction obeying a power law has been
considered in ferromagnetic spin models \cite{CaTa96}, biological networks \cite{RaGl95}, hydrodynamic interaction of active particles \cite{UCH11,GOL11b}, coupled map lattices (see \cite{lr_viana} and
references therein), and phase oscillators \cite{roger96,MAR02,lr_pre_15} in the context of {\it synchronization}. Particularly, Rogers and Wille \cite{roger96} numerically showed that a one-dimensional ring of coupled nonlinear phase oscillators with frequency mismatch 
undergoes a phase transition from a synchronized to a completely desynchronized state as the range of interaction is decreased. 

The long range power-law interaction is an ubiquitous form through which natural systems interact in physical and biological sciences. Take for example two fundamental physical interactions, namely the electromagnetic and gravitational interactions: both of them are long range in nature and obey an inverse square law of the force (i.e., $s=2$) \cite{fman}. These long range interactions are responsible for the organization of the universe on large scale (like the formation of galaxies and planetary motion) as well as on small scale (like binding of atoms and molecules in matter). In the one-dimensional Ising spin model with long range interaction of power-law type, it has been shown \cite{ising} that ferromagnetism is not possible for power-law exponent less than a critical value; similar critical power-law exponents exist in the one-dimensional spin-glass model \cite{spinglass}. In the context of biology, too, long range interaction plays an important role. Long range interaction with a specific algebraic scaling that controls the connectivity among the neurons was found in the animal brain \cite{lr_brain}. In spatial ecology, dispersal of species between different habitat patches is common. The spatial movements of most organisms are restricted and even for long-distance migrants in large networks not all the patches are likely to be accessible from a particular patch due to dispersal mortality, e.g., {\em mites} greatly suffer from dispersal mortality with increasing distance between patches \cite{BoBe09}. Moreover, to estimate the density of long-distance dispersing populations it is useful to consider that a proportion of population is distributed to the other connecting patches via a continuous geometric function so that more distant populations receive less migrants (e.g., inverse power-law) \cite{BrGuLi02}.  Interestingly, it is observed that long-distance movements of butterflies {\em Euphydryas aurinia} follow an inverse power-law \cite{FrKo07}. 

Motivated by this reasoning, in this paper we show that long range interaction with distance-dependent power-law coupling can induce various chimera patterns, like amplitude chimeras and chimera death. At the same time tuning of the power-law exponent can mediate transitions between them. Here we consider a network consisting of ecological oscillators modeled by the Rosenzweig--MacArthur system \cite{RoMa63}, which is considered as a realistic and experimentally relevant \cite{rm_expt} model in ecology. We explore and demonstrate the influence of the coupling scheme and map all the spatiotemporal behaviors including chimera states. We identify the possible transitions between the spatiotemporal patterns that arise due to the interplay of coupling strength and coupling topology characterized by the power-law exponent of the long range interaction.

\section{Coupling scheme}
We consider a ring network of $N$ nodes where the uncoupled dynamics in each node is governed by the Rosenzweig--MacArthur (RM) model \cite{RoMa63}.  
The coupled dynamics of resource and consumer are described as follows:
\begin{subequations}\label{eq1}
\begin{align}
\frac{dV_i}{dt}&=rV_i\left(1-\frac{V_i}{K}\right)-\frac{\alpha
  V_i}{V_i+B}H_i,\\ \frac{dH_i}{dt}&=H_i\left(\frac{\alpha \beta
  V_i}{V_i+B}-m\right)\nonumber\\ &
+\sigma\left(\frac{1}{\eta(s)}\sum_{p=1}^{P}\frac{H_{i-p}+H_{i+p}}{p^s}
- H_i\right),
\end{align}
\end{subequations}
where $V_i$ and $H_i$, respectively, represent the resource (or
vegetation) and consumer (or herbivore) density of the $i$-th
($i=1,\dots,N$) node, and all indices are taken modulo $N$. In the spatially extended model given by
Eqs.~(\ref{eq1}), the interaction with the neighboring nodes follows a
dispersal rate which decays with the distance $p$ between
nodes as inverse power law $1/p^{s}$ with $s \geq 0$ ($p=1,2,\cdots,
P$). Here we consider periodic boundary conditions. The spatial
dynamics are governed by the dispersal strength $\sigma$ of the
herbivore, the coupling range $P$ and the exponent of the power law
($s$); $\eta(s)=2\sum_{p=1}^{P} p^{-s}$ is the normalization constant,
with $P \leq (N-1)/2$ for odd number of nodes. It can be shown that,
even if we consider $P=(N-1)/2$, for $s\rightarrow \infty$ the
coupling reduces to a local coupling whereas for $s=0$ the coupling of
Eqs.~\eqref{eq1} represents a mean-field (global) coupling. Thus, it is
significant to note that in this {\it long range coupling}, with the
variation of the exponent $s$ one can change the nature of the coupling from
mean-field (global) coupling to local coupling via nonlocal coupling
without changing $P$.  The local (uncoupled) dynamics in each node is governed by
the following parameters: $r>0$ is the intrinsic growth rate, $K>0$ is the
carrying capacity, $\alpha>0$ is the maximum predation rate, $B>0$ is the
half saturation constant, $\beta>0$ is the conversion efficiency of
vegetation into herbivore, and $m>0$ is the mortality rate of the
herbivore.

\begin{figure}
\centering
\includegraphics[width=0.47\textwidth]{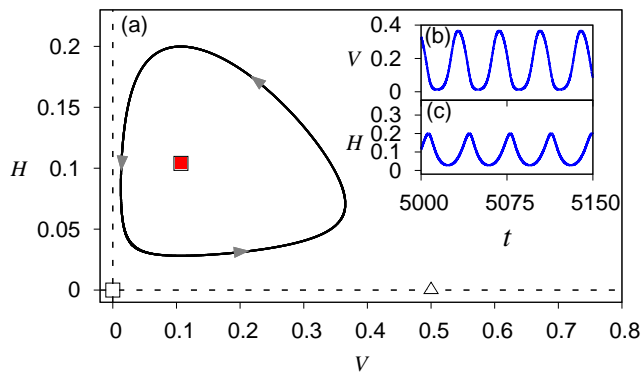}
\caption{\label{fig1b} (Color online) (a) Phase portrait of the limit cycle attractor, and
  (b), (c) time-series of the uncoupled  Rosenzweig--MacArthur model
  given by Eqs.~(\ref{eq1}) for $r= 0.5$, $K=0.5$, $\alpha=1$, $B=
  0.16$, $\beta=0.5$ and $m=0.2$, with $\sigma=0$. Symbols $\Box$ and $\Delta$
  indicate the fixed points. The red (full) square is the nontrivial fixed
  point from which the limit cycle emerges through a supercritical
  Hopf bifurcation. }
\end{figure}

\section{Results}
Before we proceed into the coupled dynamics of the network, let us
examine the local dynamics of the uncoupled system [i.e., $\sigma=0$ in Eqs.~(\ref{eq1})]. An isolated RM oscillator
has the following equilibrium points: first, $(V^*,H^*)=(0,0)$, the
eigenvalues are $(r,-m)$ and thus the equilibrium point is a saddle
point; second, $(V^*,H^*)=(K,0)$, the eigenvalues are $(-r, -m
+\alpha\beta \frac{K}{K+B})$ and the equilibrium point is either a
stable node or a saddle node, depending upon the values of the
parameters, and finally
\begin{equation}\label{eq2}
(V^*,H^*)=\left(\frac{mB}{\alpha\beta-m},
\frac{r}{\alpha}(1-\frac{mB}{K(\alpha\beta-m)})
(\frac{B\alpha\beta}{\alpha\beta-m})\right);
\end{equation}
 this nontrivial equilibrium
point is stable for parameter values satisfying the inequality
$\frac{B}{K} > \frac{(\alpha\beta-m)}{(\alpha\beta + m)}$. 
Beyond a certain $K$, this equilibrium point becomes unstable via a supercritical Hopf bifurcation and gives rise to
a stable limit cycle. A 
realistic range \cite{Mu98} of $K$ is $0.15$ to $3$, and of $m$
is $0.03$ to $0.41$. In Fig.~\ref{fig1b}, a stable limit cycle is shown
for the parameter values which are based on the experimental data reported in \cite{Mu98}.

\begin{figure}
\centering
\includegraphics[width=0.475\textwidth]{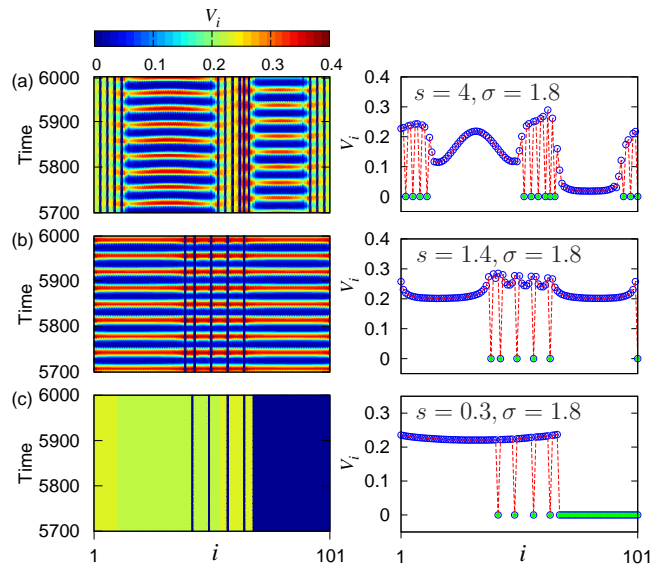}
\caption{\label{fig2} (Color online) Variation of power-law
    exponent $s$ at fixed coupling strength $\sigma$. Left panel:
  Spatiotemporal plot of vegetation $V_i$, right panel: Snapshot of $V_i$ [at
  $t=5900$, the red (dotted) lines are a guide to the eye]. Coupling range $P=50$, $N=101$, $\sigma=1.8$. 
(a) $s=4$: Amplitude chimera. (b) $s=1.4$: Chimera-like synchronized oscillation and death (CSOD). (c)
  $s=0.3$: Chimera death. In the right panel full (green) circles indicate $V_i=0$.  
Other parameters as in Fig.~\ref{fig1b}.}
\end{figure}

Next, we explore the spatiotemporal dynamics of the 
network. Our main emphasis will be to examine the effect of the
distance-dependent power-law exponent ($s$) of the long range
interaction and also to reveal the interplay of $s$ and the coupling
strength $\sigma$. Unlike other nonlocal coupling scheme we do not
explicitly vary the {\it coupling range} $P$: throughout this paper
we keep it at $P=(N-1)/2$.   Hence, we start with a globally coupled network ($s=0$) and effectively drive it to nonlocality through an increase in the inverse power-law exponent $s$.   We consider $N=101$ (i.e., $P=50$) and
integrate Eqs.~(\ref{eq1}) using the fourth-order Runge-Kutta algorithm (step
size=0.01). Figure \ref{fig2} shows the change in the spatiotemporal
dynamics at a moderate coupling strength ($\sigma=1.8$) for different values of $s$; by decreasing $s$ from panel (a) to (c)
we continuously change the coupling from near-local (a) to global (c) via nonlocal (b).
Interestingly, at $s=4$ (a) we observe a {\it two-cluster}
amplitude chimera state. In this state, the system self-organizes
into two incoherent domains separated by two coherent domains. In the
incoherent domains the oscillators show a spatially incoherent random variation in {\it
  amplitude}; at some nodes the oscillators even reach $V_i=0$. In the coherent domains the oscillators are synchronized in phase and amplitude. Next, we decrease $s$, which results in the CSOD state [Fig.~\ref{fig2}(b)]:
i.e., the network spontaneously splits into two distinct domains, in
one domain coherent oscillations occur, and in the second domain (central region) a
chimera-like coexistence of synchronized oscillations and solitary zero steady states occur in a
random spatial sequence. With further decrease in $s$, the chimera death
state emerges, which persists for lower $s$. Figure \ref{fig2}(c) shows the chimera death
 state for $s=0.3$: Here we observe two distinct subpopulations; in one subpopulation the
neighboring oscillators coherently populate either of two steady state branches: either a nontrivial steady state or the zero steady state (full green circles in the right panel). In the other subpopulation the oscillators populate the upper and
lower branch incoherently. Thus, significantly, if one goes from {\it
  near-local} to {\it near-global} coupling by simply decreasing $s$
we observe a continuous transition from amplitude chimera to chimera
death via a chimera-like (CSOD) state. We note that for $0< s
< \infty$ the coupling scheme is always nonlocal in nature, thus, the
{\it spatial connectivity} of the oscillators is always preserved in that
broad range. Further, we checked that system size ($N$) and power-law exponent $s$ have very little effect on the size of the incoherent domain in the amplitude chimera or CSOD state (not shown in the figure). As we decrease $s$, the amplitude chimera persists with an incoherent domain of almost constant size and suddenly jumps to either the chimera death state or the CSOD state (depending upon coupling strength $\sigma$).
\begin{figure} 
\centering
\includegraphics[width=0.475\textwidth]{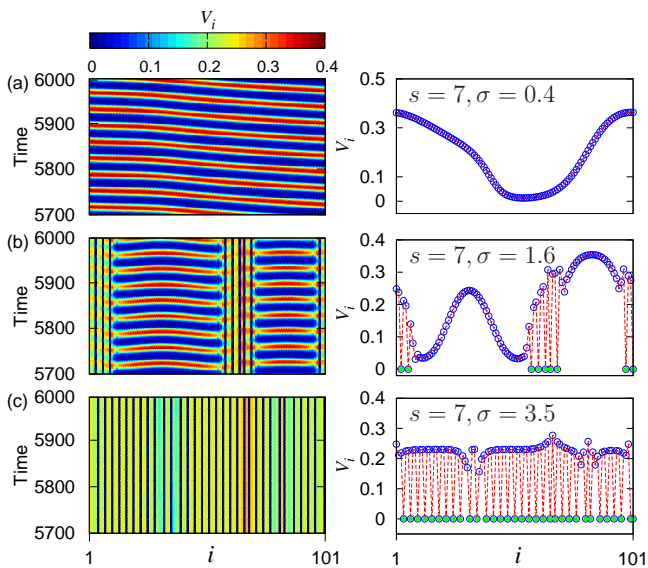}
\caption{\label{fig3} (Color online) Variation of coupling strength $\sigma$
    at fixed power-law exponent $s$. Same as Fig.~\ref{fig2} for $s=7$ and (a) $\sigma=0.4$: Synchronized oscillation,
(b) $\sigma=1.6$: Amplitude chimera. (c) $\sigma=3.5$: Multicluster oscillation death.}
\end{figure}

\begin{figure*}
\centering \includegraphics[width=0.95\textwidth]{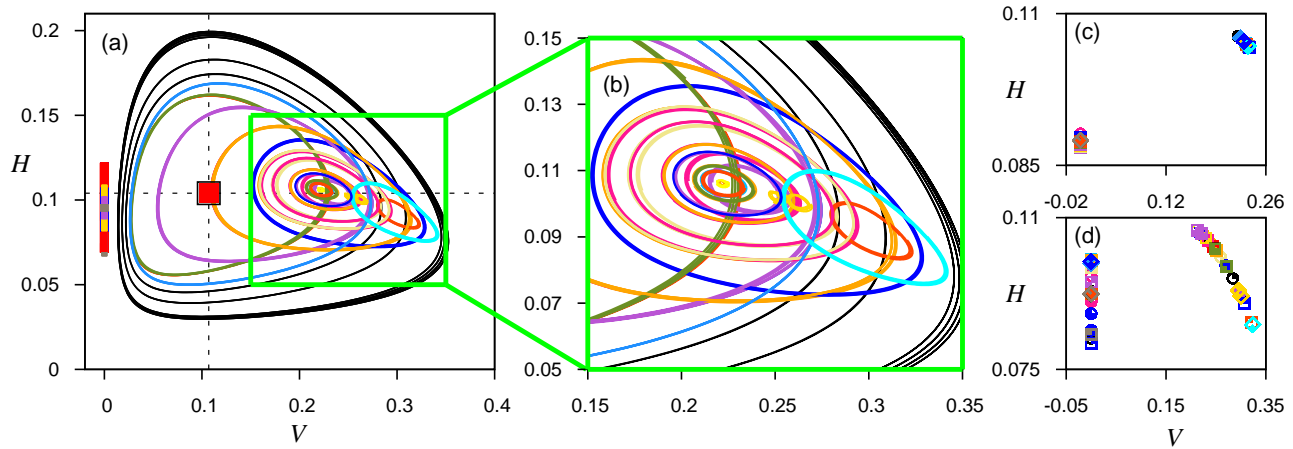}
\caption{\label{F:pp} (Color online) (a) Phase portraits (in the
  $V-H$ plane) showing the amplitude chimera of Fig.~\ref{fig2}(a). Note the shifted
  center of mass of the limit cycles of the incoherent domain. The vertical
  trajectories at $V=0$ show that in the nodes where $V$ attains a zero
  quasi-steady state, $H$ shows small amplitude oscillations. The red full square
  represents the nontrivial unstable fixed point of an
  isolated oscillator (given by Eq.~\ref{eq2}).  (b)  Blow-up of the region marked by a green box in panel (a). 
(c)--(d) Phase portraits corresponding to the chimera death shown in
    Figs.~\ref{fig2}(c) and multicluster oscillation death state of Fig.~\ref{fig3}(c), respectively. Different symbols correspond to different nodes. Other
  parameters are the same as in Fig.~\ref{fig1b}.}
\end{figure*}

Next, we fix the power-law exponent of the long range interaction at $s=7$, and vary the coupling strength $\sigma$. With an increase  in coupling strength we observe a transition from a synchronized oscillation [Fig.~\ref{fig3}(a)] to {\it multicluster oscillation death} \cite{SCH15b} [Fig.~\ref{fig3}(c)] via an amplitude chimera state [Fig.~\ref{fig3}(b)]. It is interesting to note that since we have chosen a larger value of $s$ that actually resembles a near-local coupling, we observe a traveling wave behavior in the synchronized state. Further, the structure of the death state for higher $s$ (near-local coupling) is different from that of lower $s$ (near-global coupling): in contrast to the chimera death at lower $s$ [see Fig.~\ref{fig2}(c)], here we find a multicluster oscillation death state, where the death states alternate in a regular way between upper and lower steady state branch and thus each cluster consists of only one node. Similar multicluster oscillation death states were described in the Stuart-Landau model \cite{SCH15b}.

The amplitude chimera can be best visualized in the phase space. Figure~\ref{F:pp}(a) shows the phase portrait corresponding to the amplitude chimera of Fig.~\ref{fig2}(a). Note that in the incoherent domains the centers of mass of the limit cycles associated with different nodes are shifted with respect to each other; also, they differ in their amplitudes (see also the zoom-in view in Fig.~\ref{F:pp}(b)). In contrast, in the coherent domain the oscillators share the same center of mass and have almost the same amplitude (big cycle similar to the limit cycle of the single oscillator in Fig.~\ref{fig1b}). An interesting observation can be made from Fig.~\ref{F:pp}(a): the coupled system shows isolated overlapping vertical trajectories at $V=0$. This is due to the fact that in the incoherent domain of amplitude chimeras some nodes exhibit a time-independent value $V_i=0$, however, at those nodes the variable $H_i$ shows  small amplitude oscillations. This is intuitive, because coupling is applied to the $H$ variable that permits the dispersion of herbivores ($H$) even in the nodes where $V$ reaches the zero steady state, which results in a non-zero oscillation of $H$. Figures~\ref{F:pp}(c) and \ref{F:pp}(d) show phase portraits for the chimera death state of Figs.~\ref{fig2}(c) and the multicluster oscillation death of Fig.~\ref{fig3}(c), respectively.  Here, Fig.~\ref{F:pp}(c) shows two steady states, whereas Fig.~\ref{F:pp}(d), for near-local coupling, shows two spread-out branches of steady states, as expected for near-local coupling range, see the analytical and numerical results obtained for multicluster oscillation death in \cite{SCH15b}. 

Figure~\ref{F:pp} also suggests the underlying mechanism for the occurrence of chimera patterns in the considered network. From Fig.~\ref{F:pp}(a) and \ref{F:pp}(b) one can observe the coexistence of several steady states and limit cycles of different amplitude (and center of mass), which clearly indicates the presence of multistability in the network. 

In the presence of coupling, multistability arises because the dynamical equation of each node is modified by a different coupling term depending upon the configuration of all other nodes, i.e., now every node locally adopts one of these possible states. 
This is the origin of chimera states in our present network \cite{scholl_CD,tanCD}. Note that the individual oscillators never become chaotic, rather, they remain periodic (and in a few nodes they even are in a state with $V=0$). Thus, in the incoherent domain all the oscillators are temporally periodic but spatially chaotic \cite{DMK14}. 

Further, the occurrence of steady states with $V=0$ in the incoherent region has a broad ecological significance. Since there is no
dispersal among the vegetation (V) in different patches [see Eq.~\ref{eq1}], thus, once they become extinct in a certain patch (or node) they remain so. Moreover, the carrying capacity ($K$) determines the maximum population density that an ecosystem can support. Hence, the maximum vegetation density that each patch can have is restricted by the value of $K$. Here in each patch the herbivore ($H$) survives by consuming vegetation ($V$). However, in an ecological network, herbivores may move from one patch to another. In that case, as each patch has limited vegetation density (i.e., $V_{max}=K=0.5$), in some of the patches vegetation $V$ is unable to survive (i.e., $V=0$) due to overexploitation by the herbivores. Interestingly, we note that in the nodes where vegetation $V=0$, herbivore $H\ne0$ [see Fig.~\ref{F:pp}(a)]: This is physically intuitive because dispersal occurs only in $H$. Therefore, even if $V$ in a certain patch becomes extinct, herbivores of that patch can harvest resources from the other patches present in the network and recolonize
themselves in order to avoid local extinction. But, due to the finite
coupling range (determined by $s$) they only manage to get resources from a limited number of patches which makes their oscillation amplitude small [see Fig.~\ref{F:pp}(a)]. For stronger coupling strength and larger coupling range (i.e., smaller $s$), dispersal 
through nonlocal coupling is sufficient to impose the chimera death state, i.e., both $V$ and $H$ reach stable steady states.

\begin{figure}
\centering
\includegraphics[width=0.475\textwidth]{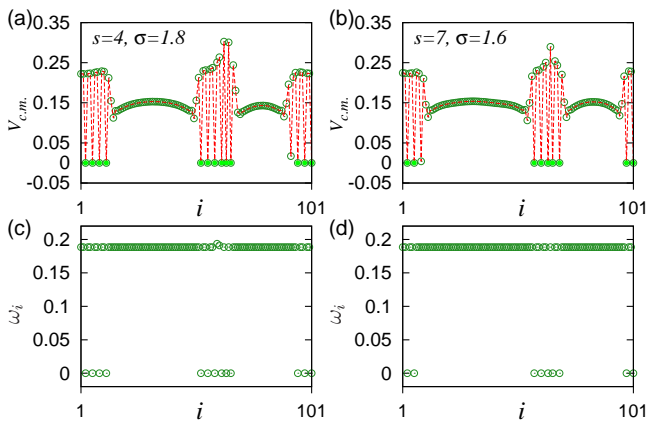}
\caption{\label{F:cm} (Color online) (a),(b) Center of mass
    $V_{c.m.}$ (Eq.~\ref{E:cm}) and (c),(d) mean phase velocity ($\omega_i$). Panels (a),(c) correspond to amplitude
    chimeras of Fig.~\ref{fig2}(a), and panels (b);(c) correspond to amplitude chimeras of Fig.~\ref{fig3}(b). Other
    parameters as in Fig.~\ref{fig1b}.}
\end{figure}

\begin{figure}
\centering
\includegraphics[width=0.44\textwidth]{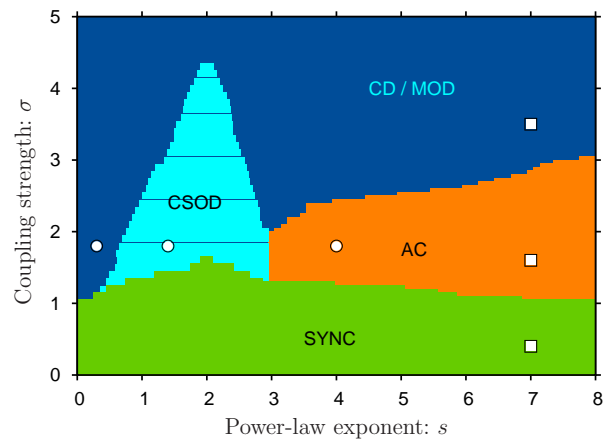}
\caption{\label{F:phase} (Color online) Phase diagram of different regimes in the
  ($s,\sigma$) plane. {\bf CSOD}: chimera-like synchronized oscillation
  and stable zero steady state (death); {\bf SYNC}: Global in-phase
  synchronized oscillation; {\bf CD/MOD}: Chimera death/multicluster oscillation death; {\bf AC}:
  Amplitude chimera. The symbols $\circ$ indicate the coupling
  parameter values used in Figs.~\ref{fig2}(a)-(c). The symbols $\Box$ correspond to Figs.~\ref{fig3}(a)-(c). Other
  parameters as in Fig.~\ref{fig1b}.}
\end{figure}

In order to provide a quantitative measure for the occurrence of amplitude chimeras we use the {\it center of mass} of each oscillator defined as in \cite{scholl_CD}:
\begin{equation}\label{E:cm}
x_{c.m}=\frac{1}{T}\int_{0}^{T}x_{i}dt,
\end{equation}
where  $x_i=V_i$ or $H_i$ of the $i$-th node. $T$ is taken sufficiently large: we use more than one thousand oscillation periods excluding the transients. The coordinate $x_{c.m.}$ actually measures the shift of a limit cycle away from the unstable fixed point from which the limit cycle has emerged (for an isolated oscillator it is the nontrivial fixed point of Eq.~(\ref{eq2})). Figures~\ref{F:cm}(a) and (b) show the plot of $V_{c.m.}$ of the amplitude chimera states of Fig.~\ref{fig2}(a) and Fig.~\ref{fig3}(b), respectively. Note that in the incoherent domains the centers of mass of the nodes show a random sequence of zero and a large positive value, whereas in the coherent domains the centers of mass exhibit a smooth profile. This spatial profile of the center of mass of oscillations is a strong characteristic of amplitude chimeras \cite{scholl_CD}. Thus, the right columns of Fig.~\ref{fig2}(a) and Fig.~\ref{fig3}(b) together with Fig.~\ref{F:cm} suggest the occurrence of amplitude chimera states.   

In Ref. \cite{scholl_CD}, pure amplitude chimeras were reported for a ring of sinusoidal Stuart-Landau oscillators, where all the oscillators have the same mean phase velocity. In contrast, in the present case we consider a model for strongly non-sinusoidal oscillators far from the Hopf bifurcation. Thus, it is interesting to investigate whether coherence-incoherence patterns like chimeras affect also the phase dynamics.  To check this we compute the mean phase velocity $\omega_i$ of each node defined as in \cite{mpv}, $\omega_i=2\pi M_i /\Delta T$, where $M_i$ is the number of oscillations during the time $\Delta T$. For our present case we use $\Delta T=10000$, and remarkably find that the mean phase velocities of all nodes are equal [Figs.~\ref{F:cm} (c), (d)]: This proves that despite the strong nonlinearity and operating point far from the Hopf bifurcation, the {\it pure} amplitude chimera is a robust and distinct chimera state: The phase part, indeed, does not play any role in amplitude chimera patterns. In Figs.~\ref{F:cm} (c) and (d), some nodes show zero mean phase velocity: those nodes actually correspond to those trajectories [vertical at $V=0$ in Fig.~\ref{F:pp}(a)] where the variable $V_i$ does not oscillate, but the variable $H_i$ does; indeed, for the $H_i$ variables we find a completely flat mean phase velocity profile (not shown here).

To explore the complete spatiotemporal dynamics of the system due to the interplay of $s$ and $\sigma$ we compute the phase diagram of regimes in the ($s,\sigma$) parameter space [see Fig.~\ref{F:phase}]. From the phase diagram it can be noticed that the choice of $s$ gives rise to three distinct transitions. For large $s$ (e.g., $s=7$, i.e., nonlocal with small coupling range) we observe transitions from the synchronized oscillation (SYNC) via amplitude chimera to multicluster oscillation death (MOD) with increasing coupling strength $\sigma$ (empty squares). In case of moderate $s$ (e.g., $0.5\lesssim s \lesssim 3$) transitions from SYNC to chimera death (CD) occur via the CSOD state. In the regime of low $s$ (i.e., near-global coupling) we observe a direct transition from SYNC to CD. Thus, it is significant that $s$ controls the overall dynamical structure of the phase diagram, and a proper choice of $s$ induces specific chimera scenarios and coherence-incoherence transitions. 

In the context of amplitude chimeras, it is important to note that, in the previous studies \cite{scholl_CD,csod,ZAK15b,LOO16,SCH16}, nonlocal coupling with a rectangular kernel has been used. In those studies it was observed that, with the variation of coupling range, which determines the degree of nonlocality, a near-local (but nonlocal) coupling supports amplitude chimeras, and as the coupling range increases chimera death emerges for most of the coupling range. In our present case, we vary the effective coupling range by controlling the power-law exponent ($s$): $s\rightarrow 0$ indicates global coupling and $s\rightarrow \infty$ means local coupling. Our results suggest that (Fig.~\ref{F:phase}) we obtain amplitude chimeras for $s \gtrsim 3$. Thus, for the distance-dependent power-law coupling, the regime of amplitude chimera becomes very broad in comparison with that for the rectangular kernel \cite{scholl_CD,csod}. However, for larger coupling strength ($\sigma$) one finds chimera death or multicluster oscillation death in a very broad zone of the parameter space (Fig.~\ref{F:phase}).

\section{Conclusion}

In this paper we have reported the occurrence of chimera patterns and associated coherence-incoherence scenarios, which are mediated by a long range interaction controlled by distance-dependent power-law scaling. Using a realistic ecological network we have shown that the variation of the power-law exponent associated with the coupling induces transitions between spatial synchrony, amplitude chimeras, and various chimera patterns like chimera death and chimeralike coexistence of synchronized oscillation and death. As the result of the interplay of coupling strength and the power law, various spatiotemporal states emerge, and we have mapped out the different regimes in the parameter space. In general, ecological networks are complex dynamical systems which are self-organized and describe species diversity, trophic (e.g., food consumption) and nontrophic (e.g., facilitation) interactions between different species and nutrients or individuals via dispersal in an ecosystem.  Moreover, species dispersal generates not only species persistence, but it also creates different types of spatio-temporal patterns, which may be associated with species invasion, colonization, or extinction in ecosystems (e.g., food webs).   In Ref.~\cite{DrStWi12} it has been predicted that species invasion attempts can be capable of breaking the synchrony in coupled ecological networks and may produce chimera states.  Our results actually support that prediction, and we are able to demonstrate the existence of various chimera patterns in an ecological network using a realistic coupling scheme.

Note that the power-law exponent associated with the coupling actually controls the overall organization of different spatiotemporal states and their mutual transitions. Thus, these results may also have applications in man-made engineering systems where by tuning the power-law exponent one can induce (or control) chimera patterns. 
As the long range interaction given by power-law scaling is very common in physics and biology, we believe that this study can be extended to other real world physical and biological systems that will enrich our understanding of chimera states.

\begin{acknowledgments}
T.B. acknowledges financial support from SERB, Department of Science and Technology (DST), India (Grant No. SB/FTP/PS-005/2013); T.B. also acknowledges the funding from SFB 910 for his visit to Institut f{\"u}r Theoretische Physik, Technische Universit\"at Berlin. P.S.D.  acknowledges financial support from SERB, DST, India [Grant No.: YSS/2014/000057]. A.Z. and E.S. acknowledge the financial support by DFG in the framework of SFB 910. A.Z. and E.S. thank Yohann Duguet for fruitful discussions.
\end{acknowledgments}

\appendix
\section{Network of Stuart-Landau oscillators}
\begin{figure}[t!]
\centering
\includegraphics[width=0.49\textwidth]{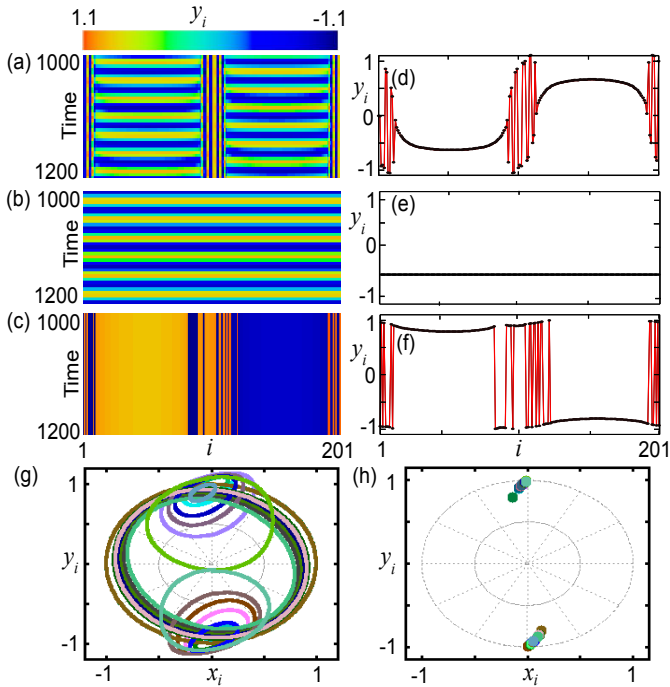}
\caption{\label{A:fig1} (Color online) Network of Stuart-Landau oscillators, coupling strength $\sigma=40$ ($\omega=2$): Spatiotemporal plots of (a) Amplitude chimera at $s=3$, (b) in-phase synchronized state at $s=2$ and (c) chimera death at $s=1$. Snapshots of (d) amplitude chimera ($s=3$), (e) in-phase synchronized state ($s=2$) and (f) chimera death ($s=1$) [red (gray) lines are a guide to the eye]. Phase portraits (g) of the amplitude chimera ($s=3$) and (h) chimera death ($s=1$).}
\end{figure}
To establish the generality of our power-law coupling scheme in inducing amplitude chimeras and other chimera patterns, like chimera death, we consider a network of Stuart-Landau (SL) oscillators, which is a generic model of nonlinear oscillators near a supercritical Hopf bifurcation. The coupled system is given by

\begin{equation}\label{sl}
\begin{split}
\dot{z_j}=&(1+i\omega-|z_j|^{2})z_j\\
&+\sigma\left(\frac{1}{\eta(s)}\sum_{p=1}^{P}\frac{\mbox{Re}(z_{j-p})+\mbox{Re}(z_{j+p})}{p^s}- \mbox{Re}(z_j)\right).
\end{split}
\end{equation}
Here $j=1\cdots N$ ($j$ is taken as modulo $N$); $z_j=x_j+iy_j$. The uncoupled oscillators have unit amplitude and an eigenfrequency $\omega$. Other parameters are as in Eq.~(\ref{eq1}). An individual SL oscillator has continuous rotational symmetry that is broken by the real-part coupling: This symmetry breaking is responsible for the appearance of chimera and chimera death \cite{scholl_CD}. Note that here also nonlocal coupling is realized in a broad range of the  power-law exponent $s$: $0<s<\infty$. It has been shown by Zakharova {\it et al}. \cite{scholl_CD} that a network of SL oscillators under nonlocal coupling exhibits amplitude chimera, in-phase synchronized states, and chimera death with proper choices of {\it coupling range} ($P$) and {\it coupling strength} ($\sigma$). Here we show that under the distance-dependent power-law coupling we can induce those states by choosing suitable values of $s$ and $\sigma$ while keeping $P$ fixed.

We consider $N=201$ identical SL oscillators with $\omega=2$. As before we take $P=\frac{(N-1)}{2}$, i.e., here $P=100$. We integrate Eq.~(\ref{sl}) using the fourth-order Runge-Kutta algorithm (step size=0.01). Spatiotemporal dynamics of the network for different $s$ with $\sigma=40$ are shown in Fig.~\ref{A:fig1} (a-c), and the corresponding snapshots are shown in Fig.~\ref{A:fig1}(d-f). For $s=3$ we observe amplitude chimeras [spatiotemporal plot in Fig.~\ref{A:fig1}(a) and snapshot in Fig.~\ref{A:fig1}(d)]; here, the incoherent domain shows a spatially random variation in amplitude (for a detailed discussion on the properties of amplitude chimeras in SL oscillators, like their transient nature, finite size effect, effect of noise, etc. see Ref.\cite{ZAK15b,LOO16,SCH16}). It is noteworthy that all the oscillators have the same phase velocity but in the incoherent domain they have disparate amplitude and center of mass. This can be seen from the phase portrait shown in Fig.~\ref{A:fig1}(g). A decrease in $s$ results in an in-phase synchronized state, which is shown in  Fig.~\ref{A:fig1}(b) [spatiotemporal plot] and Fig.~\ref{A:fig1}(e) [snapshot] for $s=2$. The chimera death pattern is observed at a relatively low $s$, i.e., in the limit of near-global coupling. Figure~\ref{A:fig1}(c) shows the spatiotemporal dynamics of the chimera death pattern for $s=1$. The corresponding snapshot and phase portrait are shown in Fig.~\ref{A:fig1}(f) and \ref{A:fig1}(h), respectively. Therefore, we conclude that as we decrease $s$ from a higher value, we observe a transition from amplitude chimera to chimera death via an in-phase synchronized state.





\begin{thebibliography}{37}%
\makeatletter
\providecommand \@ifxundefined [1]{%
 \@ifx{#1\undefined}
}%
\providecommand \@ifnum [1]{%
 \ifnum #1\expandafter \@firstoftwo
 \else \expandafter \@secondoftwo
 \fi
}%
\providecommand \@ifx [1]{%
 \ifx #1\expandafter \@firstoftwo
 \else \expandafter \@secondoftwo
 \fi
}%
\providecommand \natexlab [1]{#1}%
\providecommand \enquote  [1]{``#1''}%
\providecommand \bibnamefont  [1]{#1}%
\providecommand \bibfnamefont [1]{#1}%
\providecommand \citenamefont [1]{#1}%
\providecommand \href@noop [0]{\@secondoftwo}%
\providecommand \href [0]{\begingroup \@sanitize@url \@href}%
\providecommand \@href[1]{\@@startlink{#1}\@@href}%
\providecommand \@@href[1]{\endgroup#1\@@endlink}%
\providecommand \@sanitize@url [0]{\catcode `\\12\catcode `\$12\catcode
  `\&12\catcode `\#12\catcode `\^12\catcode `\_12\catcode `\%12\relax}%
\providecommand \@@startlink[1]{}%
\providecommand \@@endlink[0]{}%
\providecommand \url  [0]{\begingroup\@sanitize@url \@url }%
\providecommand \@url [1]{\endgroup\@href {#1}{\urlprefix }}%
\providecommand \urlprefix  [0]{URL }%
\providecommand \Eprint [0]{\href }%
\providecommand \doibase [0]{http://dx.doi.org/}%
\providecommand \selectlanguage [0]{\@gobble}%
\providecommand \bibinfo  [0]{\@secondoftwo}%
\providecommand \bibfield  [0]{\@secondoftwo}%
\providecommand \translation [1]{[#1]}%
\providecommand \BibitemOpen [0]{}%
\providecommand \bibitemStop [0]{}%
\providecommand \bibitemNoStop [0]{.\EOS\space}%
\providecommand \EOS [0]{\spacefactor3000\relax}%
\providecommand \BibitemShut  [1]{\csname bibitem#1\endcsname}%
\let\auto@bib@innerbib\@empty
\bibitem [{\citenamefont {Panaggio}\ and\ \citenamefont
  {Abrams}(2015)}]{chireview}%
  \BibitemOpen
  \bibfield  {author} {\bibinfo {author} {\bibfnamefont {M.~J.}\ \bibnamefont
  {Panaggio}}\ and\ \bibinfo {author} {\bibfnamefont {D.~M.}\ \bibnamefont
  {Abrams}},\ }\href@noop {} {\bibfield  {journal} {\bibinfo  {journal}
  {Nonlinearity}\ }\textbf {\bibinfo {volume} {28}},\ \bibinfo {pages} {R67}
  (\bibinfo {year} {2015})}\BibitemShut {NoStop}%
\bibitem [{\citenamefont {Kuramoto}\ and\ \citenamefont
  {Battogtokh}(2002)}]{KuBa02}%
  \BibitemOpen
  \bibfield  {author} {\bibinfo {author} {\bibfnamefont {Y.}~\bibnamefont
  {Kuramoto}}\ and\ \bibinfo {author} {\bibfnamefont {D.}~\bibnamefont
  {Battogtokh}},\ }\href@noop {} {\bibfield  {journal} {\bibinfo  {journal}
  {Nonlinear Phenom. Complex Syst.}\ }\textbf {\bibinfo {volume} {4}},\
  \bibinfo {pages} {380} (\bibinfo {year} {2002})}\BibitemShut {NoStop}%
\bibitem [{\citenamefont {Abrams}\ and\ \citenamefont {Strogatz}(2004)}]{st1}%
  \BibitemOpen
  \bibfield  {author} {\bibinfo {author} {\bibfnamefont {D.~M.}\ \bibnamefont
  {Abrams}}\ and\ \bibinfo {author} {\bibfnamefont {S.~H.}\ \bibnamefont
  {Strogatz}},\ }\href@noop {} {\bibfield  {journal} {\bibinfo  {journal}
  {Phys. Rev. Lett.}\ }\textbf {\bibinfo {volume} {93}},\ \bibinfo {pages}
  {174102} (\bibinfo {year} {2004})}\BibitemShut {NoStop}%
\bibitem [{\citenamefont {Abrams}\ \emph {et~al.}(2008)\citenamefont {Abrams},
  \citenamefont {Mirollo}, \citenamefont {Strogatz},\ and\ \citenamefont
  {Wiley}}]{st2}%
  \BibitemOpen
  \bibfield  {author} {\bibinfo {author} {\bibfnamefont {D.~M.}\ \bibnamefont
  {Abrams}}, \bibinfo {author} {\bibfnamefont {R.}~\bibnamefont {Mirollo}},
  \bibinfo {author} {\bibfnamefont {S.~H.}\ \bibnamefont {Strogatz}}, \ and\
  \bibinfo {author} {\bibfnamefont {D.~A.}\ \bibnamefont {Wiley}},\ }\href@noop
  {} {\bibfield  {journal} {\bibinfo  {journal} {Phys. Rev. Lett.}\ }\textbf
  {\bibinfo {volume} {101}},\ \bibinfo {pages} {084103} (\bibinfo {year}
  {2008})}\BibitemShut {NoStop}%
\bibitem [{\citenamefont {Omelchenko}\ \emph {et~al.}(2011)\citenamefont
  {Omelchenko}, \citenamefont {Maistrenko}, \citenamefont {H{\"{o}}vel},\ and\
  \citenamefont {Sch{\"{o}}ll}}]{scholl2}%
  \BibitemOpen
  \bibfield  {author} {\bibinfo {author} {\bibfnamefont {I.}~\bibnamefont
  {Omelchenko}}, \bibinfo {author} {\bibfnamefont {Y.}~\bibnamefont
  {Maistrenko}}, \bibinfo {author} {\bibfnamefont {P.}~\bibnamefont
  {H{\"{o}}vel}}, \ and\ \bibinfo {author} {\bibfnamefont {E.}~\bibnamefont
  {Sch{\"{o}}ll}},\ }\href@noop {} {\bibfield  {journal} {\bibinfo  {journal}
  {Phys. Rev. Lett.}\ }\textbf {\bibinfo {volume} {106}},\ \bibinfo {pages}
  {234102} (\bibinfo {year} {2011})}\BibitemShut {NoStop}%
\bibitem [{\citenamefont {Sethia}\ \emph {et~al.}(2013)\citenamefont {Sethia},
  \citenamefont {Sen},\ and\ \citenamefont {Johnston}}]{amc_sethia}%
  \BibitemOpen
  \bibfield  {author} {\bibinfo {author} {\bibfnamefont {G.~C.}\ \bibnamefont
  {Sethia}}, \bibinfo {author} {\bibfnamefont {A.}~\bibnamefont {Sen}}, \ and\
  \bibinfo {author} {\bibfnamefont {G.~L.}\ \bibnamefont {Johnston}},\
  }\href@noop {} {\bibfield  {journal} {\bibinfo  {journal} {Phys.Rev.E}\
  }\textbf {\bibinfo {volume} {88}},\ \bibinfo {pages} {042917} (\bibinfo
  {year} {2013})}\BibitemShut {NoStop}%
\bibitem [{\citenamefont {Zakharova}\ \emph {et~al.}(2014)\citenamefont
  {Zakharova}, \citenamefont {Kapeller},\ and\ \citenamefont
  {Sch{\"{o}}ll}}]{scholl_CD}%
  \BibitemOpen
  \bibfield  {author} {\bibinfo {author} {\bibfnamefont {A.}~\bibnamefont
  {Zakharova}}, \bibinfo {author} {\bibfnamefont {M.}~\bibnamefont {Kapeller}},
  \ and\ \bibinfo {author} {\bibfnamefont {E.}~\bibnamefont {Sch{\"{o}}ll}},\
  }\href@noop {} {\bibfield  {journal} {\bibinfo  {journal} {Phy. Rev. Lett}\
  }\textbf {\bibinfo {volume} {112}},\ \bibinfo {pages} {154101} (\bibinfo
  {year} {2014})}\BibitemShut {NoStop}%
\bibitem [{\citenamefont {Banerjee}(2015)}]{tanCD}%
  \BibitemOpen
  \bibfield  {author} {\bibinfo {author} {\bibfnamefont {T.}~\bibnamefont
  {Banerjee}},\ }\href@noop {} {\bibfield  {journal} {\bibinfo  {journal}
  {EPL}\ }\textbf {\bibinfo {volume} {110}},\ \bibinfo {pages} {60003}
  (\bibinfo {year} {2015})}\BibitemShut {NoStop}%
\bibitem [{\citenamefont {Dutta}\ and\ \citenamefont {Banerjee}(2015)}]{csod}%
  \BibitemOpen
  \bibfield  {author} {\bibinfo {author} {\bibfnamefont {P.~S.}\ \bibnamefont
  {Dutta}}\ and\ \bibinfo {author} {\bibfnamefont {T.}~\bibnamefont
  {Banerjee}},\ }\href@noop {} {\bibfield  {journal} {\bibinfo  {journal}
  {Phys. Rev. E}\ }\textbf {\bibinfo {volume} {92}},\ \bibinfo {pages} {042919}
  (\bibinfo {year} {2015})}\BibitemShut {NoStop}%
\bibitem [{\citenamefont {Hagerstrom}\ \emph {et~al.}(2012)\citenamefont
  {Hagerstrom}, \citenamefont {Murphy}, \citenamefont {Roy}, \citenamefont
  {H{\"{o}}vel}, \citenamefont {Omelchenko},\ and\ \citenamefont
  {Sch{\"{o}}ll}}]{HaMuRo12}%
  \BibitemOpen
  \bibfield  {author} {\bibinfo {author} {\bibfnamefont {A.~M.}\ \bibnamefont
  {Hagerstrom}}, \bibinfo {author} {\bibfnamefont {T.~E.}\ \bibnamefont
  {Murphy}}, \bibinfo {author} {\bibfnamefont {R.}~\bibnamefont {Roy}},
  \bibinfo {author} {\bibfnamefont {P.}~\bibnamefont {H{\"{o}}vel}}, \bibinfo
  {author} {\bibfnamefont {I.}~\bibnamefont {Omelchenko}}, \ and\ \bibinfo
  {author} {\bibfnamefont {E.}~\bibnamefont {Sch{\"{o}}ll}},\ }\href@noop {}
  {\bibfield  {journal} {\bibinfo  {journal} {Nat. Phys.}\ }\textbf {\bibinfo
  {volume} {8}},\ \bibinfo {pages} {658} (\bibinfo {year} {2012})}\BibitemShut
  {NoStop}%
\bibitem [{\citenamefont {Tinsley}\ \emph {et~al.}(2012)\citenamefont
  {Tinsley}, \citenamefont {Nkomo},\ and\ \citenamefont
  {Showalter}}]{TiNkSh12}%
  \BibitemOpen
  \bibfield  {author} {\bibinfo {author} {\bibfnamefont {M.~R.}\ \bibnamefont
  {Tinsley}}, \bibinfo {author} {\bibfnamefont {S.}~\bibnamefont {Nkomo}}, \
  and\ \bibinfo {author} {\bibfnamefont {K.}~\bibnamefont {Showalter}},\
  }\href@noop {} {\bibfield  {journal} {\bibinfo  {journal} {Nat. Phys.}\
  }\textbf {\bibinfo {volume} {8}},\ \bibinfo {pages} {662} (\bibinfo {year}
  {2012})}\BibitemShut {NoStop}%
\bibitem [{\citenamefont {Martens}\ \emph {et~al.}(2013)\citenamefont
  {Martens}, \citenamefont {Thutupalli}, \citenamefont {Fourriere},\ and\
  \citenamefont {Hallatschek}}]{MaThFo13}%
  \BibitemOpen
  \bibfield  {author} {\bibinfo {author} {\bibfnamefont {E.~A.}\ \bibnamefont
  {Martens}}, \bibinfo {author} {\bibfnamefont {S.}~\bibnamefont {Thutupalli}},
  \bibinfo {author} {\bibfnamefont {A.}~\bibnamefont {Fourriere}}, \ and\
  \bibinfo {author} {\bibfnamefont {O.}~\bibnamefont {Hallatschek}},\
  }\href@noop {} {\bibfield  {journal} {\bibinfo  {journal} {Proc. Natl. Acad.
  Sci. USA}\ }\textbf {\bibinfo {volume} {110}},\ \bibinfo {pages} {10563}
  (\bibinfo {year} {2013})}\BibitemShut {NoStop}%
\bibitem{KAP14}
T.~Kapitaniak, P.~Kuzma, J.~Wojewoda, K.~Czolczynski and Y.~Maistrenko.
\newblock Sci. Rep. {\bf 4}, 6379 (2014).
\bibitem{GAM14}
L.~V. Gambuzza, A.~Buscarino, S.~Chessari, L.~Fortuna, R.~Meucci and M.~Frasca.
\newblock Phys. Rev. E {\bf 90}, 032905 (2014).

\bibitem{LAR13}
L.~Larger, B.~Penkovsky and Y.~Maistrenko.
\newblock Phys. Rev. Lett. {\bf 111}, 054103 (2013).

\bibitem{LAR15}
L.~Larger, B.~Penkovsky and Y.~Maistrenko.
\newblock Nature Commun. {\bf 6}, 7752 (2015).

\bibitem{WIC13}
M.~Wickramasinghe and I.~Z. Kiss.
\newblock PLoS ONE {\bf 8}, e80586 (2013).

\bibitem{SCH14a}
L.~Schmidt, K.~Sch{\"o}nleber, K.~Krischer and V.~Garcia-Morales.
\newblock Chaos {\bf 24}, 013102 (2014).

\bibitem{WIC14}
M.~Wickramasinghe and I.~Z. Kiss.
\newblock Phys. Chem. Chem. Phys. {\bf 16}, 18360--18369 (2014).

\bibitem{ROS14a}
D.~P. Rosin, D.~Rontani, N.~D. Haynes, E.~Sch{\"o}ll and D.~J. Gauthier.
\newblock Phys. Rev.~E {\bf 90}, 030902(R) (2014).

\bibitem{VIK14}
E.~A. Viktorov, T.~Habruseva, S.~P. Hegarty, G.~Huyet und B.~Kelleher.
\newblock Phys. Rev. Lett. {\bf 112}, 224101 (2014).
\bibitem [{\citenamefont {Hart}\ \emph {et~al.}(2015)\citenamefont {Hart},
  \citenamefont {Bansal}, \citenamefont {Murphy},\ and\ \citenamefont
  {Roy}}]{raj_minimal}%
  \BibitemOpen
  \bibfield  {author} {\bibinfo {author} {\bibfnamefont {J.~D.}\ \bibnamefont
  {Hart}}, \bibinfo {author} {\bibfnamefont {K.}~\bibnamefont {Bansal}},
  \bibinfo {author} {\bibfnamefont {T.~E.}\ \bibnamefont {Murphy}}, \ and\
  \bibinfo {author} {\bibfnamefont {R.}~\bibnamefont {Roy}},\ }\href@noop {}
  {\bibfield  {journal} {\bibinfo  {journal} {Chaos}\ }\textbf {\bibinfo
  {volume} {26}},\ \bibinfo {pages}
  {094801} (\bibinfo {year} {2015})}\BibitemShut {NoStop}%
\bibitem [{\citenamefont {B{\"{o}}hm}\ \emph {et~al.}(2015)\citenamefont
  {B{\"{o}}hm}, \citenamefont {Zakharova}, \citenamefont {Sch{\"{o}}ll},\ and\
  \citenamefont {L{\"{u}}dge}}]{boe15}%
  \BibitemOpen
  \bibfield  {author} {\bibinfo {author} {\bibfnamefont {F.}~\bibnamefont
  {B{\"{o}}hm}}, \bibinfo {author} {\bibfnamefont {A.}~\bibnamefont
  {Zakharova}}, \bibinfo {author} {\bibfnamefont {E.}~\bibnamefont
  {Sch{\"{o}}ll}}, \ and\ \bibinfo {author} {\bibfnamefont {K.}~\bibnamefont
  {L{\"{u}}dge}},\ }\href@noop {} {\bibfield  {journal} {\bibinfo  {journal}
  {Phy. Rev. E}\ }\textbf {\bibinfo {volume} {91}},\ \bibinfo {pages}
  {040901(R)} (\bibinfo {year} {2015})}\BibitemShut {NoStop}%
\bibitem{SIE14c}
J.~Sieber, O.~Omel'chenko and M.~Wolfrum.
\newblock Phys. Rev. Lett. {\bf 112}, 054102 (2014).  
\bibitem{BIC15}
C.~Bick and E.~A. Martens.
\newblock New J.~Phys. {\bf 17}, 033030 (2015).
\bibitem{OME16}
I.~Omelchenko, O.~Omel'chenko, A.~Zakharova, M.~Wolfrum und E.~Sch{\"o}ll.
\newblock Phys. Rev. Lett. {\bf 116}, 114101 (2016).
\bibitem [{\citenamefont {Ma}\ \emph {et~al.}(2010)\citenamefont {Ma},
  \citenamefont {Wang},\ and\ \citenamefont {Liu}}]{MaWaLi10}%
  \BibitemOpen
  \bibfield  {author} {\bibinfo {author} {\bibfnamefont {R.}~\bibnamefont
  {Ma}}, \bibinfo {author} {\bibfnamefont {J.}~\bibnamefont {Wang}}, \ and\
  \bibinfo {author} {\bibfnamefont {Z.}~\bibnamefont {Liu}},\ }\href@noop {}
  {\bibfield  {journal} {\bibinfo  {journal} {Europhys. Lett.}\ }\textbf
  {\bibinfo {volume} {91}},\ \bibinfo {pages} {40006} (\bibinfo {year}
  {2010})}\BibitemShut {NoStop}%
\bibitem [{\citenamefont {Davidenko}\ \emph {et~al.}(1992)\citenamefont
  {Davidenko}, \citenamefont {Pertsov}, \citenamefont {Salomonsz},
  \citenamefont {Baxter},\ and\ \citenamefont {Jalife}}]{DaPeSa92}%
  \BibitemOpen
  \bibfield  {author} {\bibinfo {author} {\bibfnamefont {J.~M.}\ \bibnamefont
  {Davidenko}}, \bibinfo {author} {\bibfnamefont {A.~V.}\ \bibnamefont
  {Pertsov}}, \bibinfo {author} {\bibfnamefont {R.}~\bibnamefont {Salomonsz}},
  \bibinfo {author} {\bibfnamefont {W.}~\bibnamefont {Baxter}}, \ and\ \bibinfo
  {author} {\bibfnamefont {J.}~\bibnamefont {Jalife}},\ }\href@noop {}
  {\bibfield  {journal} {\bibinfo  {journal} {Nature}\ }\textbf {\bibinfo
  {volume} {355}},\ \bibinfo {pages} {349} (\bibinfo {year}
  {1992})}\BibitemShut {NoStop}%
\bibitem [{\citenamefont {Motter}\ \emph {et~al.}(2013)\citenamefont {Motter},
  \citenamefont {Myers}, \citenamefont {Anghel},\ and\ \citenamefont
  {Nishikawa}}]{grid1}%
  \BibitemOpen
  \bibfield  {author} {\bibinfo {author} {\bibfnamefont {A.~E.}\ \bibnamefont
  {Motter}}, \bibinfo {author} {\bibfnamefont {S.~A.}\ \bibnamefont {Myers}},
  \bibinfo {author} {\bibfnamefont {M.}~\bibnamefont {Anghel}}, \ and\ \bibinfo
  {author} {\bibfnamefont {T.}~\bibnamefont {Nishikawa}},\ }\href@noop {}
  {\bibfield  {journal} {\bibinfo  {journal} {Nature Phys.}\ }\textbf {\bibinfo
  {volume} {9}},\ \bibinfo {pages} {191} (\bibinfo {year} {2013})}\BibitemShut
  {NoStop}%
\bibitem{HIZ15}
J. Hizanidis, E. Panagakou, I. Omelchenko, E.~Sch{\"{o}}ll, P. H{\"{o}}vel, and A. Provata,
\newblock Phys. Rev. E {\bf 92}, 012915 (2015).
%
\bibitem [{\citenamefont {Lazarides}\ \emph {et~al.}(2015)\citenamefont
  {Lazarides}, \citenamefont {Neofotistos},\ and\ \citenamefont
  {Tsironis}}]{expt_meta}%
  \BibitemOpen
  \bibfield  {author} {\bibinfo {author} {\bibfnamefont {N.}~\bibnamefont
  {Lazarides}}, \bibinfo {author} {\bibfnamefont {G.}~\bibnamefont
  {Neofotistos}}, \ and\ \bibinfo {author} {\bibfnamefont {G.}~\bibnamefont
  {Tsironis}},\ }\href@noop {} {\bibfield  {journal} {\bibinfo  {journal}
  {Phys. Rev. B}\ }\textbf {\bibinfo {volume} {91}},\ \bibinfo {pages} {054303}
  (\bibinfo {year} {2015})}\BibitemShut {NoStop}%
\bibitem [{\citenamefont {Bastidas}\ \emph {et~al.}(2015)\citenamefont
  {Bastidas}, \citenamefont {Omelchenko}, \citenamefont {Zakharova},
  \citenamefont {Sch{\"{o}}ll},\ and\ \citenamefont {Brandes}}]{schoell_qm}%
  \BibitemOpen
  \bibfield  {author} {\bibinfo {author} {\bibfnamefont {V.~M.}\ \bibnamefont
  {Bastidas}}, \bibinfo {author} {\bibfnamefont {I.}~\bibnamefont
  {Omelchenko}}, \bibinfo {author} {\bibfnamefont {A.}~\bibnamefont
  {Zakharova}}, \bibinfo {author} {\bibfnamefont {E.}~\bibnamefont
  {Sch{\"{o}}ll}}, \ and\ \bibinfo {author} {\bibfnamefont {T.}~\bibnamefont
  {Brandes}},\ }\href@noop {} {\bibfield  {journal} {\bibinfo  {journal} {Phys.
  Rev. E}\ }\textbf {\bibinfo {volume} {92}},\ \bibinfo {pages} {062924}
  (\bibinfo {year} {2015})}\BibitemShut {NoStop}%
\bibitem{BOE15}
F.~B{\"o}hm, A.~Zakharova, E.~Sch{\"o}ll and K.~L{\"u}dge.
\newblock Phys. Rev. E {\bf 91}, 040901 (R) (2015).
\bibitem{SCH14g}
L. Schmidt and K.~Krischer.
\newblock Phys. Rev. Lett. {\bf 114}, 034101 (2015).
\bibitem{SET14}
G.~C. Sethia and A.~Sen.
\newblock Phys. Rev. Lett. {\bf 112}, 144101 (2014).
\bibitem{YEL14}
A.~Yeldesbay, A.~Pikovsky and M.~Rosenblum.
\newblock Phys. Rev. Lett. {\bf 112}, 144103 (2014).
\bibitem [{\citenamefont {Laing}(2015)}]{liang}%
  \BibitemOpen
  \bibfield  {author} {\bibinfo {author} {\bibfnamefont {C.~R.}\ \bibnamefont
  {Laing}},\ }\href@noop {} {\bibfield  {journal} {\bibinfo  {journal}
  {Phys.Rev.E}\ }\textbf {\bibinfo {volume} {92}},\ \bibinfo {pages}
  {050904(R)} (\bibinfo {year} {2015})}\BibitemShut {NoStop}%
\bibitem [{\citenamefont {Cannas}\ and\ \citenamefont
  {Tamarit}(1996)}]{CaTa96}%
  \BibitemOpen
  \bibfield  {author} {\bibinfo {author} {\bibfnamefont {S.~A.}\ \bibnamefont
  {Cannas}}\ and\ \bibinfo {author} {\bibfnamefont {F.~A.}\ \bibnamefont
  {Tamarit}},\ }\href@noop {} {\bibfield  {journal} {\bibinfo  {journal} {Phys.
  Rev. B}\ }\textbf {\bibinfo {volume} {54}},\ \bibinfo {pages} {R12661}
  (\bibinfo {year} {1996})}\BibitemShut {NoStop}%
\bibitem [{\citenamefont {Raghavachari}\ and\ \citenamefont
  {Glazier}(1995)}]{RaGl95}%
  \BibitemOpen
  \bibfield  {author} {\bibinfo {author} {\bibfnamefont {S.}~\bibnamefont
  {Raghavachari}}\ and\ \bibinfo {author} {\bibfnamefont {J.~A.}\ \bibnamefont
  {Glazier}},\ }\href@noop {} {\bibfield  {journal} {\bibinfo  {journal} {Phys.
  Rev. Lett.}\ }\textbf {\bibinfo {volume} {74}},\ \bibinfo {pages} {3297}
  (\bibinfo {year} {1995})}\BibitemShut {NoStop}%
\bibitem{UCH11}
N.~Uchida and R.~Golestanian, Phys. Rev. Lett. {\bf 106}, 058104 (2011).
\bibitem{GOL11b}
R.~Golestanian, J.~M. Yeomans, and N.~Uchida, Soft Matter {\bf 7}, 3074 (2011).
\bibitem{lr_viana}
C.~Anteneodo, S.~E.~de~S.~Pinto, A.~M.~Batista, and R.~L.~Viana.
\newblock Phys. Rev. E {\bf 68}, 045202(R) (2003).
  %
\bibitem [{\citenamefont {Rogers}\ and\ \citenamefont {Wille}(1996)}]{roger96}%
  \BibitemOpen
  \bibfield  {author} {\bibinfo {author} {\bibfnamefont {J.~L.}\ \bibnamefont
  {Rogers}}\ and\ \bibinfo {author} {\bibfnamefont {L.~T.}\ \bibnamefont
  {Wille}},\ }\href@noop {} {\bibfield  {journal} {\bibinfo  {journal} {Phys.
  Rev. E}\ }\textbf {\bibinfo {volume} {54}},\ \bibinfo {pages} {R2193}
  (\bibinfo {year} {1996})}\BibitemShut {NoStop}%
\bibitem{MAR02}
M.~Mar\'odi, F.~d'Ovidio, and T.~Vicsek, Phys. Rev. E {\bf 66}, 011109 (2002).
\bibitem [{\citenamefont {Kuo}\ and\ \citenamefont {Wu}(2015)}]{lr_pre_15}%
  \BibitemOpen
  \bibfield  {author} {\bibinfo {author} {\bibfnamefont {H.-Y.}\ \bibnamefont
  {Kuo}}\ and\ \bibinfo {author} {\bibfnamefont {K.-A.}\ \bibnamefont {Wu}},\
  }\href@noop {} {\bibfield  {journal} {\bibinfo  {journal} {Phys. Rev. E}\
  }\textbf {\bibinfo {volume} {92}},\ \bibinfo {pages} {062918} (\bibinfo
  {year} {2015})}\BibitemShut {NoStop}%
\bibitem [{\citenamefont {Feynman}\ \emph {et~al.}(2008)\citenamefont
  {Feynman}, \citenamefont {Leighton},\ and\ \citenamefont {Sands}}]{fman}%
  \BibitemOpen
  \bibfield  {author} {\bibinfo {author} {\bibfnamefont {R.~P.}\ \bibnamefont
  {Feynman}}, \bibinfo {author} {\bibfnamefont {R.~B.}\ \bibnamefont
  {Leighton}}, \ and\ \bibinfo {author} {\bibfnamefont {M.}~\bibnamefont
  {Sands}},\ }\href@noop {} {\emph {\bibinfo {title} {The Feynman Lectures on
  Physics: Vol. 1--3}}}\ (\bibinfo  {publisher} {Narosa},\ \bibinfo {address}
  {New Delhi, India},\ \bibinfo {year} {2008})\BibitemShut {NoStop}%
\bibitem [{\citenamefont {Aizenman}\ \emph {et~al.}(1988)\citenamefont
  {Aizenman}, \citenamefont {Chayes}, \citenamefont {Chayes},\ and\
  \citenamefont {Neuman}}]{ising}%
  \BibitemOpen
  \bibfield  {author} {\bibinfo {author} {\bibfnamefont {M.}~\bibnamefont
  {Aizenman}}, \bibinfo {author} {\bibfnamefont {J.~T.}\ \bibnamefont
  {Chayes}}, \bibinfo {author} {\bibfnamefont {L.}~\bibnamefont {Chayes}}, \
  and\ \bibinfo {author} {\bibfnamefont {C.~M.}\ \bibnamefont {Neuman}},\
  }\href@noop {} {\bibfield  {journal} {\bibinfo  {journal} {J. Stat. Phys.}\
  }\textbf {\bibinfo {volume} {50}},\ \bibinfo {pages} {1} (\bibinfo {year}
  {1988})}\BibitemShut {NoStop}%
\bibitem [{\citenamefont {Kotliar}\ \emph {et~al.}(1983)\citenamefont
  {Kotliar}, \citenamefont {Anderson},\ and\ \citenamefont
  {Stein}}]{spinglass}%
  \BibitemOpen
  \bibfield  {author} {\bibinfo {author} {\bibfnamefont {G.}~\bibnamefont
  {Kotliar}}, \bibinfo {author} {\bibfnamefont {P.~W.}\ \bibnamefont
  {Anderson}}, \ and\ \bibinfo {author} {\bibfnamefont {D.~L.}\ \bibnamefont
  {Stein}},\ }\href@noop {} {\bibfield  {journal} {\bibinfo  {journal} {Phys.
  Rev. B}\ }\textbf {\bibinfo {volume} {27}},\ \bibinfo {pages} {602} (\bibinfo
  {year} {1983})}\BibitemShut {NoStop}%
\bibitem [{\citenamefont {Szaro}\ and\ \citenamefont
  {Tompkins}(1987)}]{lr_brain}%
  \BibitemOpen
  \bibfield  {author} {\bibinfo {author} {\bibfnamefont {B.~G.}\ \bibnamefont
  {Szaro}}\ and\ \bibinfo {author} {\bibfnamefont {R.}~\bibnamefont
  {Tompkins}},\ }\href@noop {} {\bibfield  {journal} {\bibinfo  {journal} {J.
  Comp. Neurol}\ }\textbf {\bibinfo {volume} {258}},\ \bibinfo {pages} {304}
  (\bibinfo {year} {1987})}\BibitemShut {NoStop}%
\bibitem [{\citenamefont {Bowler}\ and\ \citenamefont {Benton}(2009)}]{BoBe09}%
  \BibitemOpen
  \bibfield  {author} {\bibinfo {author} {\bibfnamefont {D.~E.}\ \bibnamefont
  {Bowler}}\ and\ \bibinfo {author} {\bibfnamefont {T.~G.}\ \bibnamefont
  {Benton}},\ }\href@noop {} {\bibfield  {journal} {\bibinfo  {journal}
  {Oikos}\ }\textbf {\bibinfo {volume} {118}},\ \bibinfo {pages} {403}
  (\bibinfo {year} {2009})}\BibitemShut {NoStop}%
\bibitem [{\citenamefont {Brown}\ \emph {et~al.}(2002)\citenamefont {Brown},
  \citenamefont {Gupta}, \citenamefont {Li}, \citenamefont {Milne},
  \citenamefont {Restrepo},\ and\ \citenamefont {West}}]{BrGuLi02}%
  \BibitemOpen
  \bibfield  {author} {\bibinfo {author} {\bibfnamefont {J.~H.}\ \bibnamefont
  {Brown}}, \bibinfo {author} {\bibfnamefont {V.~K.}\ \bibnamefont {Gupta}},
  \bibinfo {author} {\bibfnamefont {B.}~\bibnamefont {Li}}, \bibinfo {author}
  {\bibfnamefont {B.~T.}\ \bibnamefont {Milne}}, \bibinfo {author}
  {\bibfnamefont {C.}~\bibnamefont {Restrepo}}, \ and\ \bibinfo {author}
  {\bibfnamefont {G.~B.}\ \bibnamefont {West}},\ }\href@noop {} {\bibfield
  {journal} {\bibinfo  {journal} {Philosophical Transactions of the Royal
  Society B: Biological Sciences}\ }\textbf {\bibinfo {volume} {357}},\
  \bibinfo {pages} {619} (\bibinfo {year} {2002})}\BibitemShut {NoStop}%
\bibitem [{\citenamefont {Frica}\ and\ \citenamefont
  {Konvicka}(2007)}]{FrKo07}%
  \BibitemOpen
  \bibfield  {author} {\bibinfo {author} {\bibfnamefont {Z.}~\bibnamefont
  {Frica}}\ and\ \bibinfo {author} {\bibfnamefont {M.}~\bibnamefont
  {Konvicka}},\ }\href@noop {} {\bibfield  {journal} {\bibinfo  {journal}
  {Basic and Applied Ecology}\ }\textbf {\bibinfo {volume} {8}},\ \bibinfo
  {pages} {377} (\bibinfo {year} {2007})}\BibitemShut {NoStop}%
\bibitem [{\citenamefont {Rosenzweig}\ and\ \citenamefont
  {MacArthur}(1963)}]{RoMa63}%
  \BibitemOpen
  \bibfield  {author} {\bibinfo {author} {\bibfnamefont {M.~L.}\ \bibnamefont
  {Rosenzweig}}\ and\ \bibinfo {author} {\bibfnamefont {R.~H.}\ \bibnamefont
  {MacArthur}},\ }\href@noop {} {\bibfield  {journal} {\bibinfo  {journal} {The
  American Naturalist}\ }\textbf {\bibinfo {volume} {97}},\ \bibinfo {pages}
  {209} (\bibinfo {year} {1963})}\BibitemShut {NoStop}%
\bibitem [{\citenamefont {Fussmann}\ \emph {et~al.}(2000)\citenamefont
  {Fussmann}, \citenamefont {Ellner}, \citenamefont {Shertzer},\ and\
  \citenamefont {Jr.}}]{rm_expt}%
  \BibitemOpen
  \bibfield  {author} {\bibinfo {author} {\bibfnamefont {G.}~\bibnamefont
  {Fussmann}}, \bibinfo {author} {\bibfnamefont {S.~P.}\ \bibnamefont
  {Ellner}}, \bibinfo {author} {\bibfnamefont {K.~W.}\ \bibnamefont
  {Shertzer}}, \ and\ \bibinfo {author} {\bibfnamefont {N.~G.}\ \bibnamefont
  {Hairston Jr.}},\ }\href@noop {} {\bibfield  {journal} {\bibinfo  {journal} {Science}\
  }\textbf {\bibinfo {volume} {290}},\ \bibinfo {pages} {1358} (\bibinfo {year}
  {2000})}\BibitemShut {NoStop}%
\bibitem [{\citenamefont {Murdoch}\ \emph {et~al.}(1998)\citenamefont
  {Murdoch}, \citenamefont {Nisbet}, \citenamefont {McCauley}, \citenamefont
  {deRoos},\ and\ \citenamefont {Gurney}}]{Mu98}%
  \BibitemOpen
  \bibfield  {author} {\bibinfo {author} {\bibfnamefont {W.~W.}\ \bibnamefont
  {Murdoch}}, \bibinfo {author} {\bibfnamefont {R.~M.}\ \bibnamefont {Nisbet}},
  \bibinfo {author} {\bibfnamefont {E.}~\bibnamefont {McCauley}}, \bibinfo
  {author} {\bibfnamefont {A.~M.}\ \bibnamefont {deRoos}}, \ and\ \bibinfo
  {author} {\bibfnamefont {W.~S.~C.}\ \bibnamefont {Gurney}},\ }\href@noop {}
  {\bibfield  {journal} {\bibinfo  {journal} {Ecology}\ }\textbf {\bibinfo
  {volume} {79}},\ \bibinfo {pages} {1339} (\bibinfo {year}
  {1998})}\BibitemShut {NoStop}%
\bibitem{SCH15b}
I.~Schneider, M.~Kapeller, S.~Loos, A.~Zakharova, B.~Fiedler, and E.~Sch{\"{o}}ll.
\newblock Phys. Rev. E {\bf 92}, 052915 (2015).
\bibitem{DMK14}
D. Dudkowski, Y. Maistrenko, and T. Kapitaniak.
\newblock Phys. Rev. E {\bf 90}, 032920 (2014).
\bibitem [{\citenamefont {Omelchenko}\ \emph {et~al.}(2013)\citenamefont
  {Omelchenko}, \citenamefont {Omelchenko}, \citenamefont {H{\"{o}}vel},\ and\
  \citenamefont {Sch{\"{o}}ll}}]{mpv}%
  \BibitemOpen
  \bibfield  {author} {\bibinfo {author} {\bibfnamefont {I.}~\bibnamefont
  {Omelchenko}}, \bibinfo {author} {\bibfnamefont {O.}~\bibnamefont
  {Omelchenko}}, \bibinfo {author} {\bibfnamefont {P.}~\bibnamefont
  {H{\"{o}}vel}}, \ and\ \bibinfo {author} {\bibfnamefont {E.}~\bibnamefont
  {Sch{\"{o}}ll}},\ }\href@noop {} {\bibfield  {journal} {\bibinfo  {journal}
  {Phys. Rev. Lett.}\ }\textbf {\bibinfo {volume} {110}},\ \bibinfo {pages}
  {224101} (\bibinfo {year} {2013})}\BibitemShut {NoStop}%
\bibitem{ZAK15b}
A.~Zakharova, M.~Kapeller, and E.~Sch{\"{o}}ll.
\newblock J. Phys. Conf. Series {\bf 727}, 012018 (2016), arXiv 1503.03371.
\bibitem{LOO16}
S.~Loos, J.~C.~Claussen, E.~Sch{\"{o}}ll, and A.~Zakharova.
\newblock Phys. Rev. E {\bf 93}, 012209 (2016).
\bibitem{SCH16}
E.~Sch{\"{o}}ll, S.~H.~L.~Klapp, and P.~H{\"o}vel (eds.): {\em Control of self-organizing nonlinear systems} (Springer, Berlin, 2016).
\bibitem{DrStWi12}
J.~A.~Drake, P.~Staelens, and D.~Wieczynski, in {\em Encyclopedia of Theoretical Ecology} edited by A.~Hastings and
L.~Gross (University of California Press, 2012), pp.~60--63.

\end{thebibliography}
\providecommand{\noopsort}[1]{}\providecommand{\singleletter}[1]{#1}%

\end{document}